# First-principles equation of state and phase stability for the Ni-Al system under high pressures

H. Y. Geng[a, c, *], N. X. Chen[a, b] and M. H. F. Sluiter[d]

[a] *Department of Physics, Tsinghua University, Beijing 100084, China*

[b] *Institute for Applied Physics, University of Science and Technology, Beijing 100083, China*

[c] *Laboratory for Shock Wave and Detonation Physics Research, Southwest Institute of Fluid Physics, P. O. Box 919-102, Mianyang Sichuan 621900, China*

[d] *Institute for Materials Research, Tohoku University, Sendai, 980-8577 Japan*

**Abstract**: The equation of state (EOS) of alloys at high pressures is generalized with the cluster expansion method. It is shown that this provides a more accurate description. The low temperature EOSs of Ni-Al alloys on FCC and BCC lattices are obtained with density functional calculations, and the results are in good agreement with experiments. The merits of the generalized EOS model are confirmed by comparison with the mixing model. In addition, the FCC phase diagram of the Ni-Al system is calculated by cluster variation method (CVM) with both spin-polarized and non-spin-polarized effective cluster interactions (ECI). The influence of magnetic energy on the phase stability is analyzed. A long-standing discrepancy between *ab initio* formation enthalpies and experimental data is addressed by defining a better reference state. This aids both evaluation of an *ab initio* phase diagram and understanding the thermodynamic behaviors of alloys and compounds. For the first time the high-pressure behavior of order-disorder transition is investigated by *ab initio* calculations. It is found that order-disorder temperatures follow the Simon melting equation. This may be instructive for experimental and theoretical research on the effect of an order-disorder transition on shock Hugoniots.

## I. INTRODUCTION

In recent years, the first-principles theory of alloy phase stability of simple crystal structures and their superstructures has much advanced, and the study of complex phases, where several inequivalent sites exist in the unit cell, has gradually attracted the interest of theoretical investigations.[1-8] However, there remain significant issues in the study of phase stability of simple crystal structures.[9-11] Notably, the effect of pressure on the thermodynamic properties and the phase diagram (PD) of alloys have been investigated in few works only.[12-14] One of the authors (MS) has found by *ab initio* calculations that the Al-Li system is not affected significantly by hydrostatic compression, except for some very minor effects, such as the reduced Li solubility in the Al-rich fcc solid solution.[13] However, the pressure in that computation is limited to 5.4GPa, and the conclusion is for one specific system only. The most important issues of high-pressure physics of alloys, e.g., the equation of state (EOS), have not been studied yet. Progress in the physics of the earth's interior indicates that there are many nontrivial pressure-temperature and pressure-composition phase diagrams for mantle minerals. A similar situation for alloys with complex structure can be expected. The present work on alloys and compounds at high pressures, their equations of state and phase stability is undertaken to better understand the pressure behavior of alloys. The Ni-Al system was selected because it is the basis of Ni-based superalloys. It is necessary to point out that although the thermodynamics of Ni-Al binary system have been studied in great detail (including both experiments and theoretical calculations),[11, 15-20] almost all of these works apply to zero pressure and high pressure behavior remains unknown.

[*] Corresponding author. Department of Physics, Tsinghua University, Beijing 100084, China.
*E-mail address*: genghy02@mails.tsinghua.edu.cn

The theory of the EOS for alloys and compounds remains rather undeveloped; the prevalent model being the mixing model or the so-called volume-addition model.[21-23] The basic assumption of this model is that the volume of alloys or compounds under pressure is given by the summation of equilibrium volumes of its constituents,

$$V(P)=\sum_i n_i v_i(P)\,, \tag{1}$$

where $v_i(P)$ is the equilibrium volume of $i$th component at pressure $P$ and $n_i$ the concentration. The internal energy is then given by

$$E(P)=\sum_i n_i \varepsilon_i(P)\,, \tag{2}$$

and the enthalpy is as

$$H=\sum_i n_i(\varepsilon_i(v_i)+Pv_i)=\sum_i n_i H_i(v_i)\,. \tag{3}$$

This model assumes that thermodynamic quantities are just the arithmetic average of each constituent, and more subtle details, say, the structure-dependence of these quantities, are ignored.

Here, we suggest a more general EOS model based on the cluster expansion (CE) method. The EOS of Ni-Al alloys are investigated by density functional calculations at zero temperature and the generalized CE EOS model in the tetrahedron approximation is compared with the mixing model. Spin-polarization effects on phase stability in the Ni-Al system are explored and are shown to have partly obscured the fair assessment of *ab initio* results. Finally, the order-disorder transition temperature dependence on pressure in FCC Ni-Al alloys is investigated for first time.

## II. THEORETICAL MODEL

### A. Generalization of EOS model for alloys

For generalizing the mixing EOS model, the cluster expansion method (CEM) [24-27] is a natural choice for the mixing model in fact corresponds to the point approximation of CEM, where it is always assumed that interactions are short-ranged in order to guarantee the convergence.

The internal energy and pressure in trinomial EOS[28] are separated as $E=E_x+E_v+E_e$ and $P=P_x+P_v+P_e$, where subscripts *x*, *v* and *e* refer to the contribution at zero Kelvin, the thermal contribution from lattice vibrations and that of thermal electrons, respectively. Ionization due to temperature and compression is beyond the scope of this work and ignored. With CEM, one can write the (free) energy terms as functions of correlation functions as[24]

$$E_x(V) = \sum_n v_n(V)\xi_n \tag{4}$$

for the zero temperature part of internal energy and

$$F_v(V,T) = \sum_n w_n(V,T)\xi_n \tag{5}$$

for the free energy of thermal vibrations,[29] where $\xi$ is the cluster correlation function as defined in Eq.(10) in Ref.12. As for the electronic free energy, instead of the simple free-electrons approximation (which is almost configurational independent),[21-22] it is better to use integration involving the configurational electronic density of state $n_\sigma(E)$:

$$F_e(\sigma,T) = \int^{\mu(T)} n_\sigma(E)\left[Ef(E) + k_B T\left[f(E)\ln f(E) + (1-f(E)\ln(1-f(E)))\right]\right]dE\,, \tag{6}$$

where $f(E)$ is the Fermi-Dirac distribution. Then, CEM is employed to obtain the electronic free energy for any configuration,

$$F_e = \sum_n \lambda_n(V,T)\xi_n\,. \tag{7}$$

The convergence of this expansion is heuristic and further confirmation is needed.

Pressure can be formulated analogously by $P_x = -\partial E_x / \partial V$ and $P_T = -(\partial F_T / \partial V)_T$:

$$P_x(V) = -\sum_n v_n'(V)\xi_n\,, \tag{8}$$

$$P_v(V,T) = -\sum_n w_n'(V,T)\xi_n\,, \tag{9}$$

$$P_e(V,T) = -\sum_n \lambda_n'(V,T)\xi_n\,, \tag{10}$$

where the prime indicates the derivative with respect to volume. Eqs.(8-10) compose the generalized EOS model which has the capability to account for the effects of order-disorder transitions in alloys. Provided that effective cluster interactions (ECI) $v_n$, $w_n$ and $\lambda_n$ are known, either from *ab initio* calculations or from fitting to experimental data, the thermodynamic properties and equilibrium state can be computed readily by the cluster variation method (CVM).[30] It is evident now that the mixing model is indeed the single point approximation of CE EOS model as pointed out before. In this paper, we will focus mainly on the zero temperature compressions and vibrational[31-32] and thermal electronic effects all are neglected.

## B. Calculation methodology

For we do not aim to model magnetic transitions,[33] the magnetic cohesive energies as well as enthalpies of Ni-Al system can be approximated by simple spin-polarized calculations. Total energies of FCC-based superstructures for Ni-Al system (FCC, $L1_0$, $L1_2$ and $DO_{22}$), as well as those based on BCC lattice (BCC, B2, B32 and $DO_3$), are computed within the generalized gradient approximation[34-36] by CASTEP (CAmbridge Serial Total Energy Package)[37-38] with a range of lattice parameters. Both spin-polarized and non-polarized results are calculated in order to evaluate the influence of magnetic energy on phase diagram. All calculations are performed using ultrasoft pseudopotentials.[39] The cutoff kinetic energy for planewaves in the expansion of the wave functions is set as 540eV. Integrations in reciprocal space are performed in the first Brillouin zone using a grid with a maximal interval of 0.03/Å generated by Monkhorst-Pack[40] scheme. The energy tolerance for self-consistent field (SCF) convergence is $2\times10^{-6}$eV/atom for all calculations. This setting gives a precision of 0.2*m*eV/atom to the convergence of the total energy for FCC Al.

Cohesive energies at different lattice parameters are extracted from the total energies by subtracting the spin-polarized energies of isolated atoms. Then, they are employed to evaluate the CE EOS at zero-Kelvin and the formation enthalpies for CVM[13, 30, 41-45] calculations according to

$$\Delta H^{\alpha}_{form}(P) = H^{\alpha}(P) - c^{\alpha}_{A} H^{A-\alpha}(P) - (1-c^{\alpha}_{A}) H^{B-\alpha}(P)\,, \tag{11}$$

where superscript $\alpha$ refers to superstructure, and $c^{\alpha}_{A}$ the concentration of species *A* in $\alpha$ phase. *P* is hydrostatic pressure and enthalpy is defined as

$$H^{\alpha}(P) = E^{\alpha}_{coh}[V(P)] + PV(P)\,. \tag{12}$$

The volume $V$ is determined directly by solving $P = -\partial E_{coh} / \partial V$ in this work, implying the effects of heat expansion have been neglected. After formation enthalpies $\Delta H_{form}$ of a set of superstructures have been worked out, the effective cluster interaction (ECI) $v_n(P)$ for cluster $n$ at pressure $P$ can be obtained readily by means of a Connolly-Williams procedure[24]

$$v_n(P) = \sum_{\alpha} \Delta H_{form}^{\alpha}(P)(\xi_n^{\alpha})^{-1} . \tag{13}$$

This set of ECIs is appropriate for phase stability calculations. However, it is improper for EOS computations since cohesive energies and their pressure-dependence of pure elements have been omitted. A set of ECIs containing more information needed for EOS, while may be less accurate for phase stability studies, can be derived analogously by

$$\bar{v}_n(V) = \sum_{\alpha} E_{coh}^{\alpha}(V)(\xi_n^{\alpha})^{-1} . \tag{14}$$

Here $\bar{v}_n$ corresponds to the contribution of cluster $n$ to cohesive energy. Eqs.(13) and (14) can be solved using a singular value decomposition procedure. Then, the EOS of any phase can be calculated based on its cohesive energy curve

$$E_{coh}^{\alpha}(V) = \sum_{n=1}^{n_{\max}} \bar{v}_n(V)\xi_n^{\alpha} . \tag{15}$$

Merits of Eq.(15) lie on its capability of providing accurate EOS for alloys (in particular solid solutions) that is difficult by direct *ab initio* methods. The phase equilibria at finite temperatures are determined with the Gibbs free energy by CVM

$$G^{\alpha} = H^{\alpha} - TS^{\alpha} . \tag{16}$$

In present work, only tetrahedron approximation is used because we focus mainly on the trends and variations of phase boundaries and transition temperatures rather than the precise phase diagram and tetrahedron is enough for this purpose.[46]

## III. RESULTS AND DISCUSSION

### A. EOS at zero temperature

Calculated cohesive energies, equilibrium lattice parameters and bulk moduli are listed in tables I to IV. Experimental and other theoretical results are also included for comparisons. The superscripts in tables refer to the corresponding reference papers. Both spin-polarized and non-polarized results are presented simultaneously to evaluate the influence of local moments on weak magnetic Ni-Al alloys. The cohesive energies for a range of atomic volume are calculated and shown in figures 1-2. For elemental Al, the spin-polarized and non-polarized cohesive energy curves are identical within a large range of volume, which is different from elemental Ni (Fig.2). The excess energies due to spin-polarization of valence electrons are about -0.5(-1)eV for FCC Al(Ni) at a lattice parameter of 15Å. These values are comparable to cohesive energies of Ni-Al alloys at ambient pressure and accurate cohesive energies can be obtained only when referenced them to spin-polarized isolated atoms.

For nonmagnetic phase of B2 and FCC Al, The calculated equilibrium lattice parameters and bulk moduli are in good agreement with experimental data[47-52] (better than previous calculations[16, 53-55]). Our computed lattice parameters are slightly larger than other calculations systematically. It is owing to the GGA (GGS) approximation, which always overcorrects the deficiencies of LDA and leads to an underbinding. The influence of spin-polarization of electrons are limited to Ni-rich side with concentration of Al below 0.5(0.25) for FCC(BCC) based phases. Spin-polarized equilibrium lattice parameters of magnetic phase (FCC Ni and $L1_2$ $Ni_3Al$) are better than non-polarized ones by comparing with experimental data (partly for this reason, following discussions at zero temperature are all based on spin-polarized calculations if without special statements). The calculated bulk modulus of FCC Ni, both spin-polarized and non-polarized, however, are larger than experiment measurements. This is expected since DFT calculations always overestimate the cohesive energy and consequently the bulk modulus for transition metals.

Based on above calculations, the cold EOS (spin-polarized) of Ni-Al alloys is computed readily. Shown in figure 3 is the pressure vs. compression ratio curves, whose feature of concentration and structure dependences is evident. It demonstrates the mixing model is inappropriate for ordered states. The curves of B2 and $L1_0$ phases are almost identical, and those of $DO_{22}$ and $L1_2$ are close very well within the studied pressure range. In particular, a detailed comparison of these curves with experimental[52] and the mixing model results is given in figure 4 for stoichiometric NiAl, where FCC+FCC (BCC+BCC) curve is derived from FCC(BCC) elemental phases only by mixing model. B32 phase seems better than the stable B2 phase by compared with experimental data. However, both of them are within the measurement error bar. The curves of bulk modulus vs compression ratio are also presented in figure 5. One can see both the bulk modulus and its gradient with respect to volume of non-polarized FCC Ni are larger than the spin-polarized one. The structure dependence of bulk modulus is also evident.

The EOS of Ni-Al alloys can be generally calculated using ECIs obtained by Eq.(14). For the purpose to justify the CE EOS model, a stable phase of stoichiometric $L1_2$ $Ni_3Al$ is considered. The ECIs for pressure are shown in figure 6, which are derived from those for cohesive energies by $p_n = -\partial \bar{v}_n(V) / \partial V$ (for bulk modulus, $b_n = V \partial^2 \bar{v}_n(V) / \partial V^2$ is applied analogously). Under tetrahedron approximation, *n* takes the value from zero to four, corresponding to null cluster, point, nearest neighbor (NN) pair, NN triangle and NN tetrahedron, respectively. Limited by the used parent cluster and superstructures, the coefficients for clusters of point and NN pair are identical (this degeneracy is lifted when lager cluster and more superstructures are used).

Convergence of cluster expansion is demonstrated by the decrease of ECIs' magnitude by ten times successively. Figure 7 shows the comparisons of bulk modulus, cohesive energy and pressure between results of mixing model and the CE EOS model, respectively. Subscript FP refers to first-principles calculations. Obviously, CE EOS is much better than the mixing model, although the latter also provided a relative precise approximation to the first principles results. Peaks in the figure correspond to the zero points of first-principle cohesive energy, pressure and bulk modulus and indicate the requirement of larger parent cluster for more accurate EOS.

**B. Phase stability**

The spin-polarized formation enthalpies of Ni-Al system as functions of pressure are plotted in figure 8 with pressure up to 400GPa. A structural transition from FCC to BCC takes place at about 260GPa for Al. It is in agreement with previous calculations except for a more stable phase, HCP, which is not considered here, presents at 220-300GPa at low temperature.[56] The stability of all ordered phases are strengthened by pressure, while $DO_3$ is more notable comparing with $DO_{22}$ phase. The comparison of our calculated formation enthalpies at zero pressure with experimental data[57-59] and previous calculations[16, 53] is shown in figure 9. Both spin-polarized and non-polarized results are included. It is clear that the former is much better by comparing with the experimental data. The latter, however, shallower than Pasturel's results[16] and in good agreement with Watson's calculations.[53] All theoretical calculations predict the same order of stabilities for studied phases. The discrepancy between the theoretical results and experimental data at Al-rich side is due to that more stable phases, $DO_{20}$ ($NiAl_3$) and $D5_{13}$ ($Ni_2Al_3$), in this composition range are not considered in this work. It is necessary to point out that the experiment data of Oelsen[60] is excluded for their measurements were not rigorous.[19]

The phase stability of Ni-Al system at finite temperature is computed with CVM and Eq.(13) is employed to derive the corresponding ECIs. To evaluate the influence of magnetic energy on phase stability partly, FCC phase diagrams (PD) are produced by both spin-polarized and non-polarized ECIs. Figure 10 shows the low temperature part of this PD. It is surprised that the spin-polarized and non-polarized PDs are almost identical. The only discernable distinction is $L1_2$-FCC boundaries at Ni-rich side shown in the inset. This is unusual for the two sets of ECIs are quite different. A completely different situation presents for high temperature part, however (see Fig.11). The reason for this lies on that the Gibbs free energy depends on both ECIs and entropy. Its variation with respect to small changes of ECIs $v_n \to v_n + \delta v_n$ is simply as

$$\delta G^{\alpha} \approx \sum_n \delta v_n \xi_n^{\alpha} - \frac{T}{2} \sum_n \frac{\partial^2 S^{\alpha}}{\left(\partial \xi_n^{\alpha}\right)^2} (\Delta \xi_n^{\alpha})^2 . \tag{17}$$

Here the condition $\partial G^{\alpha} / \partial \xi_n^{\alpha} = 0$ is used, and $\Delta \xi_n^{\alpha}$ are variations of correlation functions due to the changes of ECIs via the procedure of minimizing Gibbs energy. One concludes from figures 10 and 11 that the contribution of the first term in Eq.(17) is small, while the second term is magnified by temperature $T$ and becomes dominant at high temperatures. The distinct phase boundaries at Ni-rich side (Fig.11) are just the responsibility of this term, indicating the precision requirement of ECIs for reliable Gibbs free energy and phase diagram calculations at high temperatures.

We also find from figure 11 that the spin-polarized ECIs produced a wrong high temperature PD for Ni-Al alloys. The order-disorder transition temperature $T_c$ of $L1_2$ $Ni_3Al$-FCC is too low to be true. In fact, it is still too low even volume relaxation effects are included. This crushes Carlsson *et al*'s hope[20] to improve the first-principles $T_c$ by including magnetic energy. It is reasonable because the range of temperature here is much higher than the Curie temperatures of Ni-Al alloys and the magnetic interactions should have been vanished. Thus the proper ECIs for this region should be the non-polarized one. Actually, the non-polarized PD is in agreement with previous calculations,[16, 20] and an improvement of $T_c$ about 100K is acquired when no volume relaxation effects included. The relaxed $T_c$ is about 2500K with an improvement of 300K compared with previous calculations,[16] counting roughly 15% of the extrapolated experimental $T_c$. This result can be improved further by employing larger parent clusters, including local lattice distortions and vibrational entropies.[31]

Nevertheless, it is inconsistent between experimental formation enthalpies and the phase diagram. The former prefers to the spin-polarized ECIs whereas the latter prefers to non-polarized one. The situation becomes worse when formation enthalpies measured at different temperatures are taken into account. It seems the formation enthalpy of Ni-Al alloys is scatted and intractable.[53] However, if dividing the measured formation enthalpies into two sets according to whether they are measured below or above the Curie temperature of Ni, one may find those measured at low temperatures (commonly at room temperature) prefers to spin-polarized results, while the other set prefers to non-polarized one. Obviously the excess spin-polarized energy of Ni is the key for this problem. In view of almost all ordered phases of Ni-Al system are nonmagnetic at room temperature except FCC Ni, it is convenient to shift the reference state from magnetic Ni (used in measurements) to nonmagnetic state for these data. This is done using the spin-polarized and non-polarized cohesive energies of FCC Ni listed in Tables I and II. The low temperature experimental formation enthalpy of $Ni_3Al$ is then reevaluated from –37.3[58] (-35[53])KJ/mol to –53.8(-51.5)KJ/mol, which is in good agreement with our non-polarized result –47.0KJ/mol, Pasturel's –48.36KJ/mol,[16] and high temperature measurement of –47KJ/mol.[53] That of NiAl (B2) is also reevaluated from -58.8KJ/mol[57] to -69.79KJ/mol, by comparison with our non-polarized -67.3KJ/mol, Pasturel's –75.6KJ/mol,[16] and high temperature measurement of –67KJ/mol.[53] It is evident now that the discrepancy between the experiment data and Pasturel's calculations is mainly due to LDA approximation they used, which has been corrected in this work by GGA instead.

**C. Simon equation for order-disorder transition temperature**

It is interesting to investigate the variation of order-disorder transition temperatures $T_c$ of L$1_2$ $Ni_3Al$ and L$1_0$ NiAl phases with pressures. Here only cold pressure is taken into account up to 130GPa for simplicity, which is determined by non-polarized cohesive energy curves and no vibrational contributions are included. The $T_c$ of L$1_0$ phase is lower than that of L$1_2$ only within a narrow range of pressure and has a larger gradient (see Fig.12). It is worth to point out that $T_c$ satisfies perfectly the Simon's melting equation,[61] which is a semi-empirical law for melting at high pressures. The reason for this may lie in that both order-disorder transformations (L$1_2$-FCC and L$1_0$-FCC) and melting are first order. We know the phase boundary of a first-order transition must obey the Clausius-Clapeyron relation

$$\frac{dP}{dT} = \frac{\Delta S}{\Delta V} = \frac{\Delta H}{T\Delta V}. \qquad (18)$$

On the other hand, Simon equation has a form of

$$\frac{P - P_0}{a} = \left(\frac{T}{T_0}\right)^c - 1. \qquad (19)$$

One can then obtain a relation for the latent heat, pressure and difference of volume for order-disorder transition as $\frac{\Delta H}{c\Delta V} = a + P$. The parameters *a* and *c* are 40.249GPa and 3.546 for L$1_2$ $Ni_3Al$ and 21.472GPa and 2.935 for L$1_0$ NiAl, respectively.

The significance of this relation is that it would ignite the interest to investigate the high-pressure thermodynamic behaviors of alloys, in particular the influence of order-disorder transition on shock Hugoniots. A heuristic question is for B2-BCC transition. It is second order and what kind of relation will be followed by its $T_c$? Is it still in Simon form or not? All of these are still open for answers.

## IV. CONCULUSION

In conclusion, the mixing model for high pressure EOS of alloys is generalized to CE EOS model with the cluster expansion method. It is shown that this provides a more accurate description of ordered state due to its feature of structure dependence. The low temperature EOSs of Ni-Al alloys that based on FCC/BCC lattice are calculated by first-principles method and a good agreement with experiment data is obtained. The CE EOS model is confirmed by comparison with the mixing model in tetrahedron approximation. We also provide the formation enthalpies of studied structures up to 400GPa in order to analyze the variation of phase stability as functions of pressure. The FCC phase diagram of Ni-Al system is calculated by CVM with both spin-polarized and non-polarized ECIs to evaluate the influence of magnetic energy. By defining a more sound reference state, the low temperature experimental formation enthalpies are reevaluated and the results matched very well with our first-principles calculations, previous *ab initio* results and high temperature measurements simultaneously, addressing the long standing discrepancy of the formation enthalpies for Ni-Al system. For the first time the high-pressure behavior of order-disorder transition is investigated by *ab initio* calculations. It is found that order-disorder temperatures follow the Simon melting equation. This may be instructive for experimental and theoretical research on the effect of an order-disorder transition on shock Hugoniots.

## V. ACKNOWLEDGMENTS

This work was supported by the National Advanced Materials Committee of China. And the authors gratefully acknowledge the financial support from 973 Project in China under Grant No. G2000067101.

**Figure Captions:**

Fig.1. Spin-polarized cohesive energies vs atomic volume for some BCC and FCC structrures.

Fig.2. Comparison of cohesive energies of BCC Ni with spin-polarized and non-polarized FCC Ni.

Fig.3. *Ab initio* pressure-compression ratio curves for Ni-Al alloys based on FCC and BCC lattices. Notice the structure-dependences.

Fig.4. Comparison of calculated EOS with experimental (Otto *et al.*) and the mixing model results for NiAl.

Fig.5. Calculated spin-polarized bulk moduli as functions of compression ratio.

Fig.6. Cluster expansion coefficients for pressure in tetrahedron approximation.

Fig.7. Comparisons of cluster expansion EOS with mixing model referenced to first-principles results in terms of cohesive energy, pressure and bulk modulus, respectively.

Fig.8. Formation enthalpies as functions of pressure up to 400GPa. Notice the strengthening of the stability of $DO_3$, B2 and BCC Al phases.

Fig.9. Calculated formation enthalpies at zero pressure compared with experimental and previous theoretical results. The convex hull pertaining to spin-polarized (non-polarized) ground states is marked with a solid (dotted) line.

Fig.10. FCC phase diagram of Ni-Al system at low temperature region.

Fig.11. FCC phase diagram of Ni-Al system at high temperature region. Notice the Ni-rich part, where spin-polarized ECIs produced wrong phase boundaries.

Fig.12. Calculated order-disorder transition temperature as functions of pressure by comparison with Simon equation.

**Table Captions:**

Table I. Spin-polarized total energies for FCC superstructures at 0GPa.

Table II. Non-polarized total energies for FCC superstructures at 0GPa.

Table III. Spin-polarized total energies for BCC superstructures at 0GPa.

Table IV. Non-polarized total energies for BCC superstructures at 0GPa.

Table I.

| Structure (spin-polarized) | $c_{\mathrm{Al}}$ | $E_{\mathrm{coh}}$ (eV/atom) | $a$ (Å) | $a_{other}$ (Å) | $B$ (GPa) | $B_{other}$ (GPa) |
|---|---|---|---|---|---|---|
| fcc | 0.0 | -4.873 | 3.510 | 3.52[50] | 215.6 | 187.6[47] |
| | | | | 3.450[16] | | |
| $DO_{22}$ | 0.25 | -4.825 | 3.557 | 3.54[54] | 190.5 | |
| | | | | 3.538[16] | | |
| $L1_2$ | 0.25 | -4.873 | 3.547 | 3.567[51] | 194.5 | 186[47] |
| | | | | 3.55[54] | | |
| | | | | 3.532[16] | | |
| $L1_0$ | 0.5 | -4.624 | 3.651 | 3.569[53] | 159.4 | |
| | | | | 3.613[16] | | |
| $DO_{22}$ | 0.75 | -4.008 | 3.845 | 3.777[53] | 112.5 | |
| | | | | 3.781[16] | | |
| $L1_2$ | 0.75 | -4.009 | 3.839 | 3.802[16] | 111.1 | |
| fcc | 1.0 | -3.498 | 4.052 | 4.05[50] | 78.6 | 79.4[47] |
| | | | | 3.984[53] | | |

Table II.

| Structure (non-polarized) | $c_{\mathrm{Al}}$ | $E_{\mathrm{coh}}$ (eV/atom) | $a$ (Å) | $a_{other}$ (Å) | $B$ (GPa) |
|---|---|---|---|---|---|
| fcc | 0.0 | -4.645 | 3.488 | 3.52[50] | 227.4 |
| | | | | 3.450[16] | |
| $DO_{22}$ | 0.25 | -4.806 | 3.553 | 3.54[54] | 195.5 |
| | | | | 3.538[16] | |
| $L1_2$ | 0.25 | -4.845 | 3.545 | 3.567[51] | 198.3 |
| | | | | 3.55[54] | |
| | | | | 3.532[16] | |
| $L1_0$ | 0.5 | -4.624 | 3.651 | 3.569[53] | 159.0 |
| | | | | 3.613[16] | |
| $DO_{22}$ | 0.75 | -4.008 | 3.845 | 3.777[53] | 111.1 |
| | | | | 3.781[16] | |
| $L1_2$ | 0.75 | 4.009 | 3.839 | 3.802[16] | 111.4 |
| fcc | 1.0 | -3.498 | 4.052 | 4.05[50] | 79.2 |
| | | | | 3.984[53] | |

Table III.

| Structure (spin-polarized) | $c_{\mathrm{Al}}$ | $E_{\mathrm{coh}}$ (eV/atom) | $a$ (Å) | $a_{other}$ (Å) | $B$ (GPa) | $B_{other}$ (GPa) |
|---|---|---|---|---|---|---|
| bcc | 0.0 | -4.731 | 2.794 | 2.745[16] | 210.0 | |
| *DO*$_3$ | 0.25 | -4.788 | 2.825 | 2.755[53] | 188.4 | |
| | | | | 2.789[16] | | |
| *B*2 | 0.5 | -4.769 | 2.882 | 2.886[49] | 162.1 | 166[48] |
| | | | | 2.833[53] | | 156±3[52] |
| | | | | 2.864[16] | | 186[55] |
| *B*32 | 0.5 | -4.438 | 2.914 | 2.871[16] | 151.2 | |
| *DO*$_3$ | 0.75 | -3.879 | 3.056 | 3.003[16] | 105.0 | |
| bcc | 1.0 | -3.403 | 3.240 | 3.177[16] | 71.3 | |

Table IV.

| Structure (non-polarized) | $c_{\mathrm{Al}}$ | $E_{\mathrm{coh}}$ (eV/atom) | $a$ (Å) | $a_{other}$ (Å) |
|---|---|---|---|---|
| bcc | 0.0 | -4.592 | 2.774 | 2.745[16] |
| *DO*$_3$ | 0.25 | -4.787 | 2.821 | 2.755[53] |
| | | | | 2.789[16] |
| *B*2 | 0.5 | -4.769 | 2.882 | 2.886[49] |
| | | | | 2.833[53] |
| | | | | 2.864[16] |
| *B*32 | 0.5 | -4.438 | 2.914 | 2.871[16] |
| *DO*$_3$ | 0.75 | -3.879 | 3.056 | 3.003[16] |
| bcc | 1.0 | -3.403 | 3.240 | 3.177[16] |

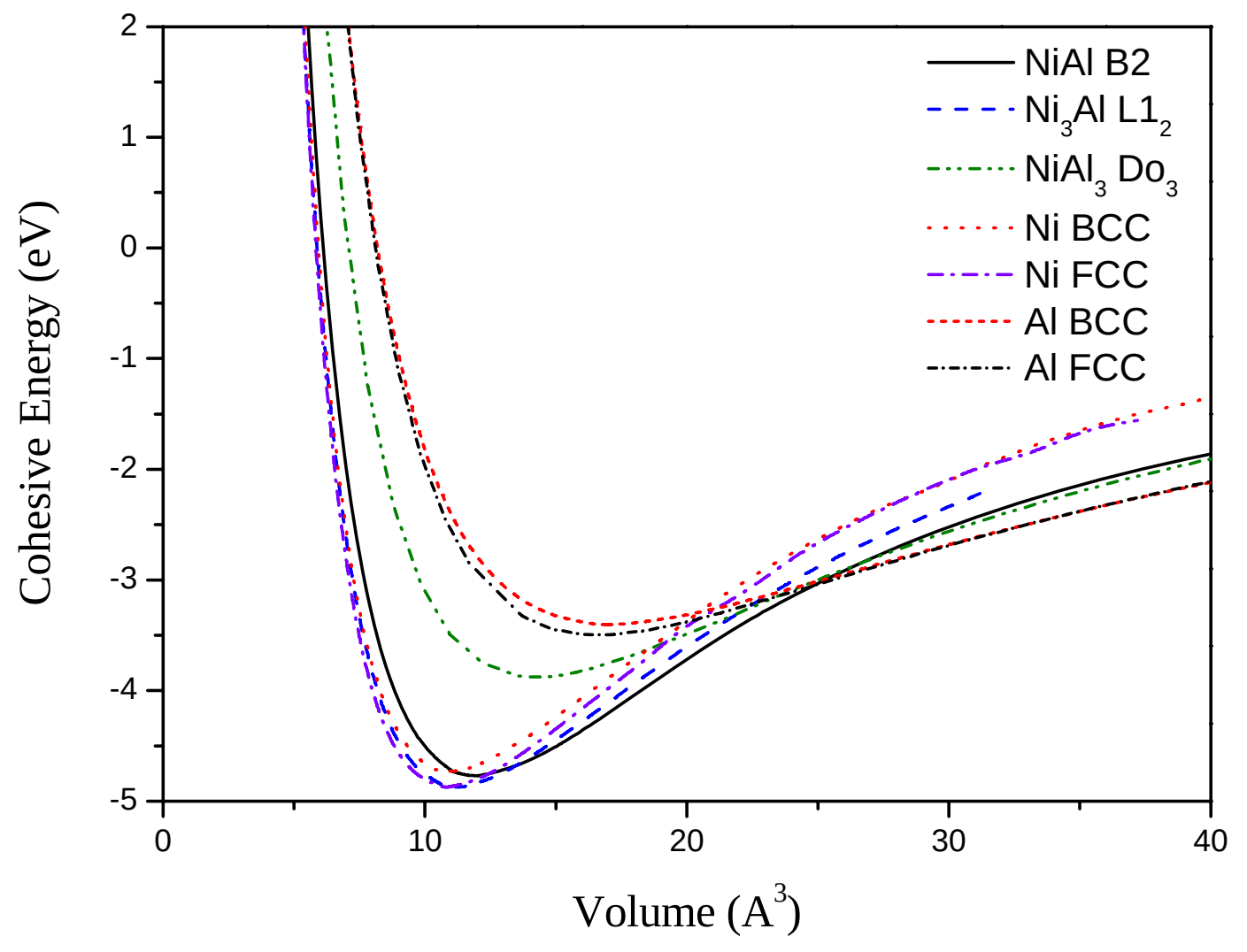

Fig. 1.

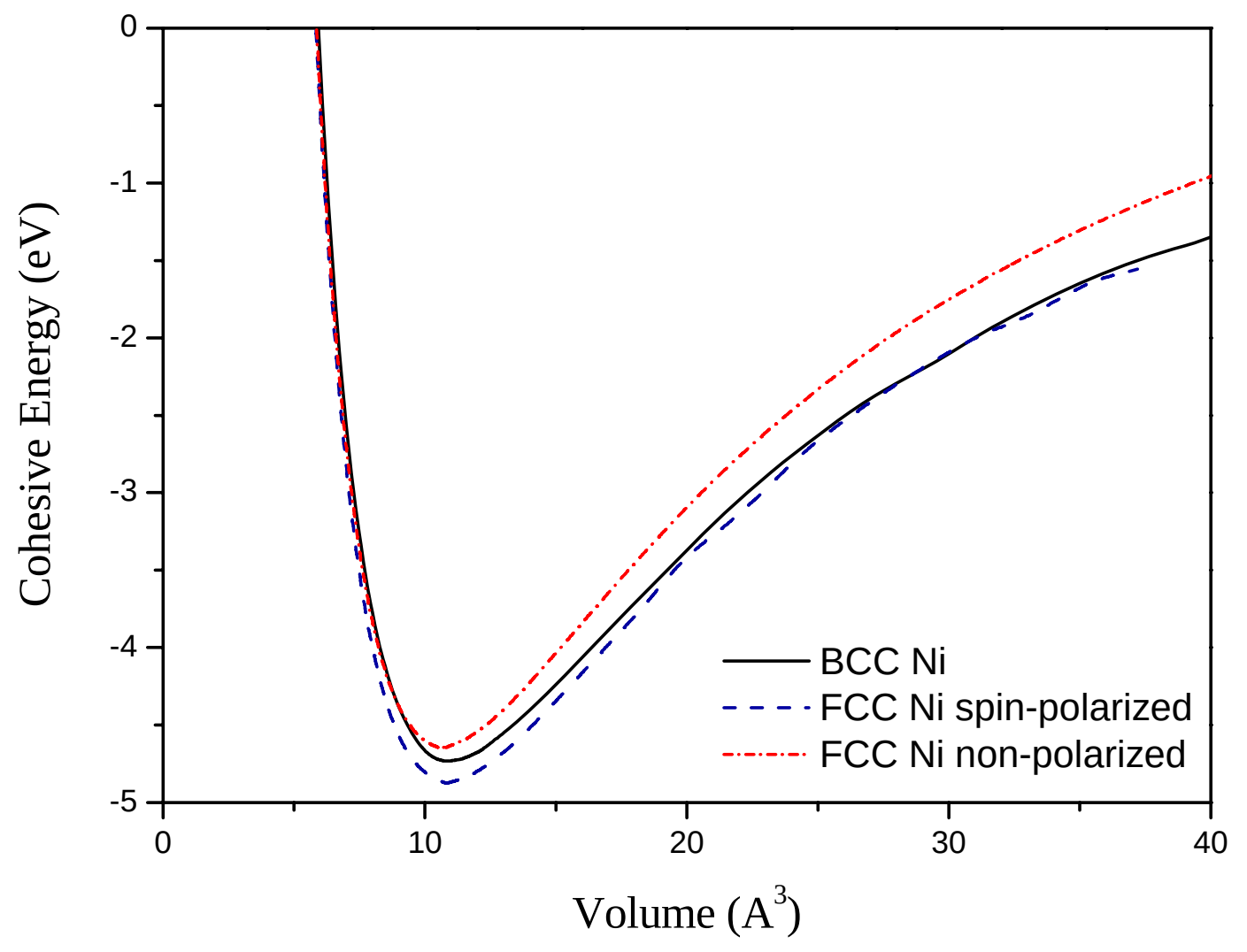

Fig. 2.

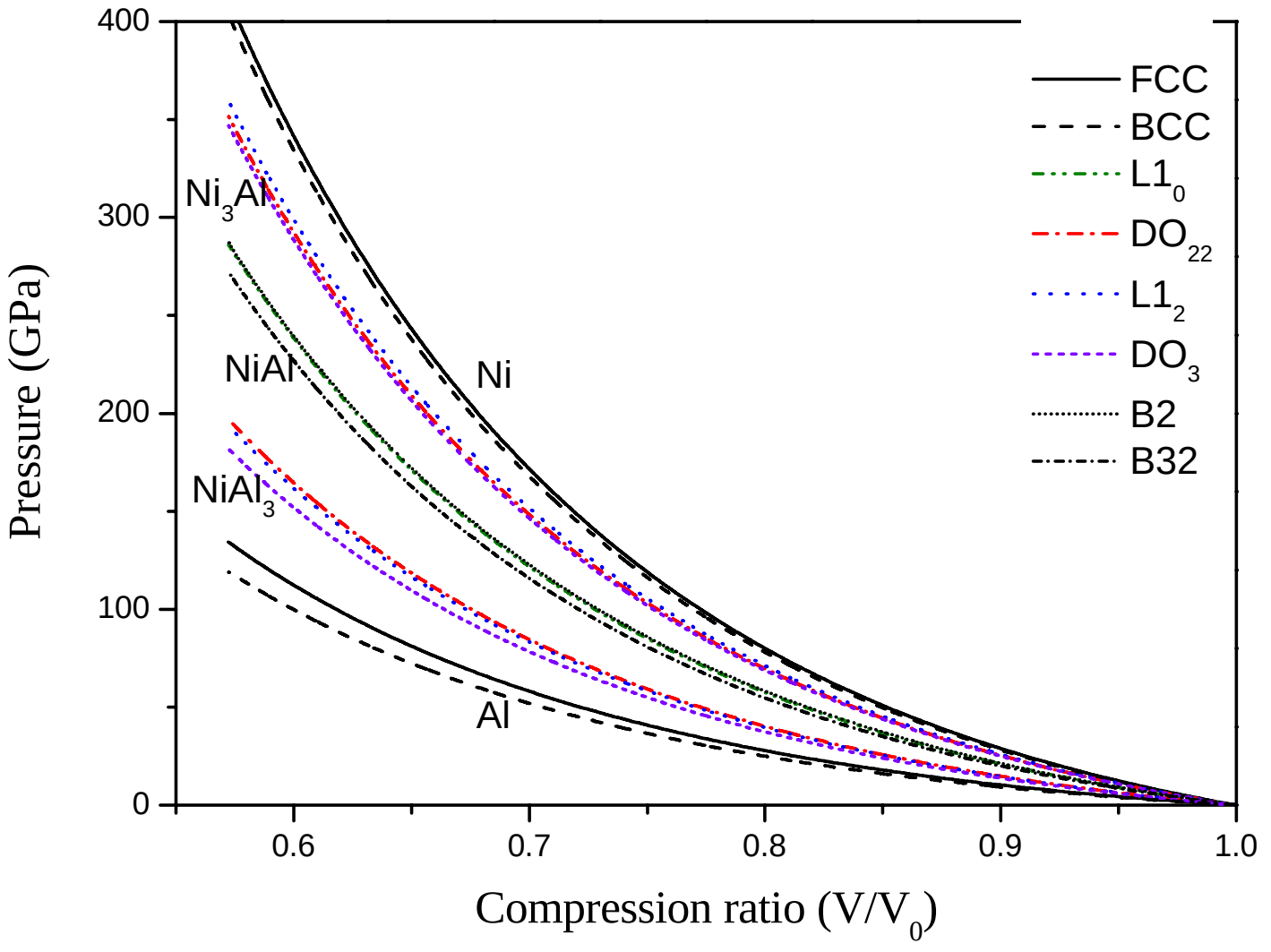

Fig.3.

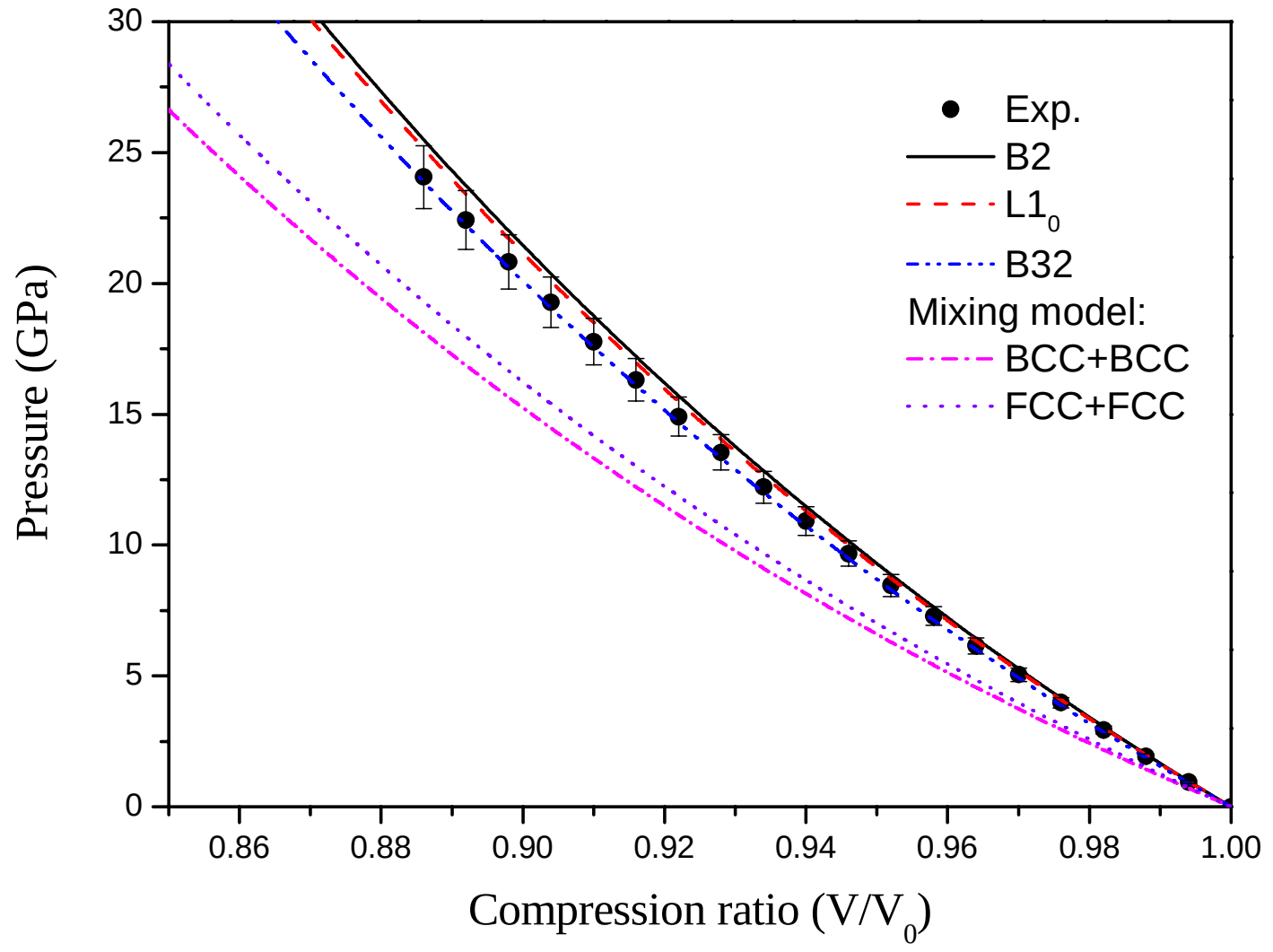

Fig.4.

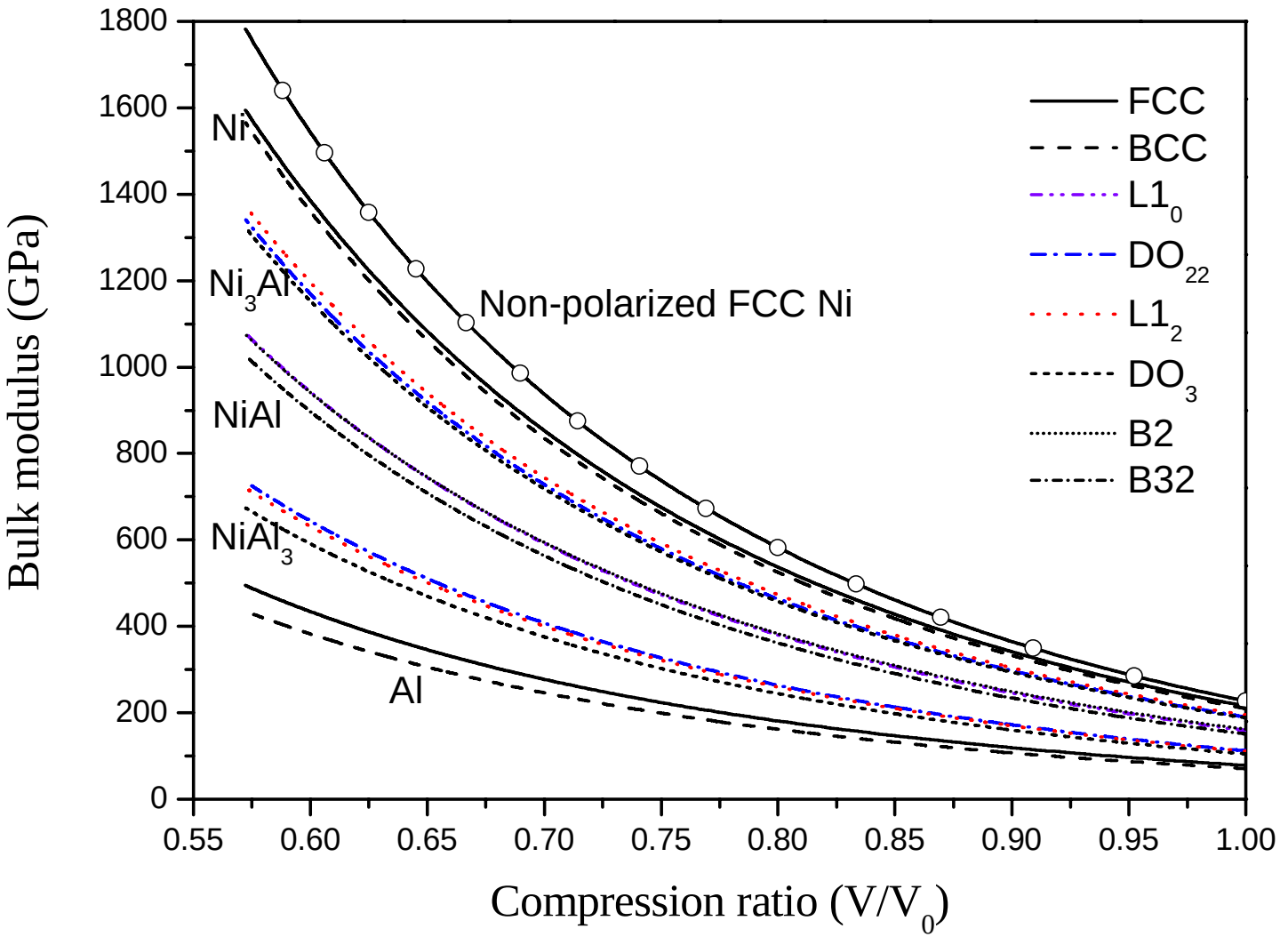

Fig. 5.

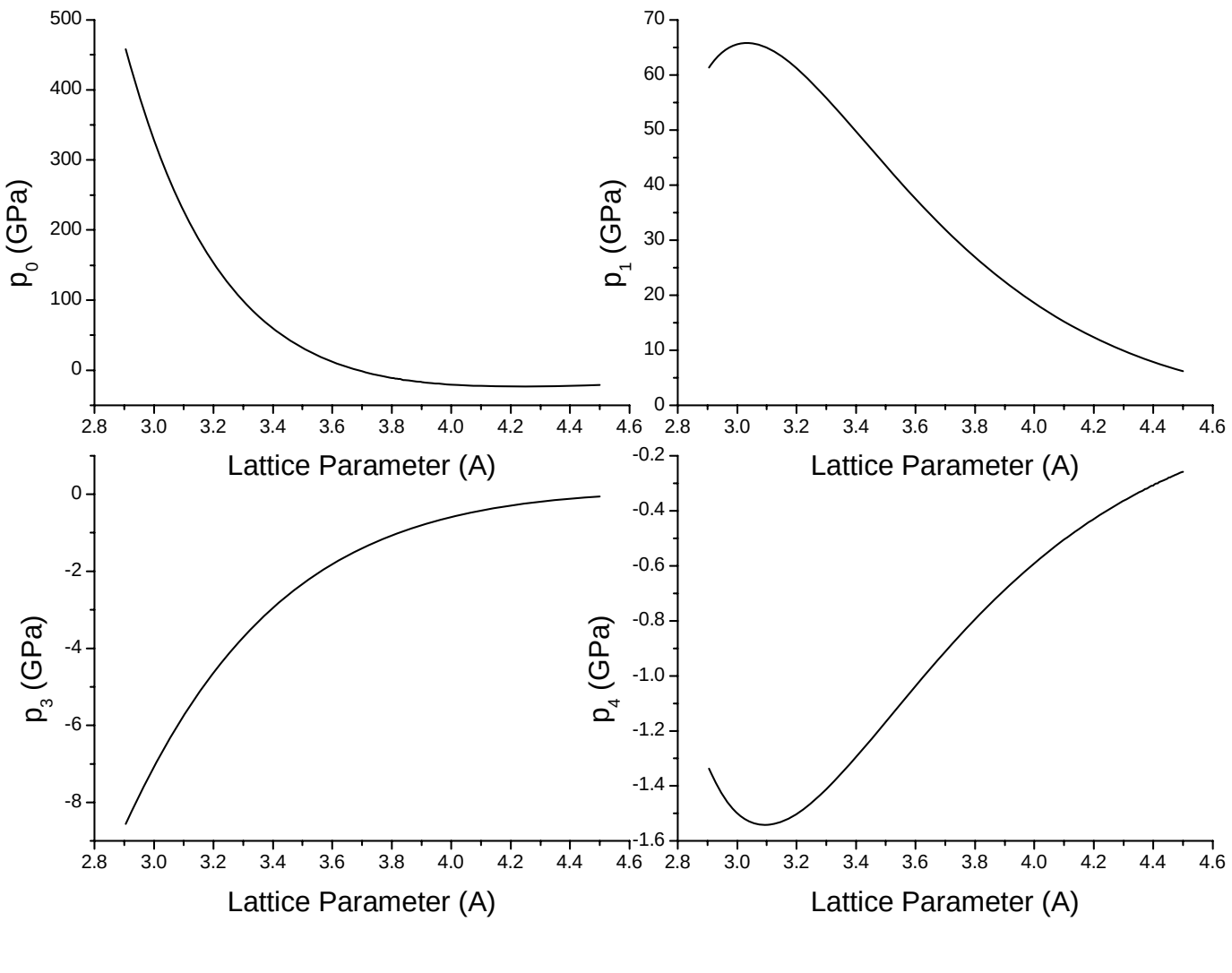
p0 (GPa)
p1 (GPa)
p3 (GPa)
p4 (GPa)
Lattice Parameter (A)

Fig. 6.

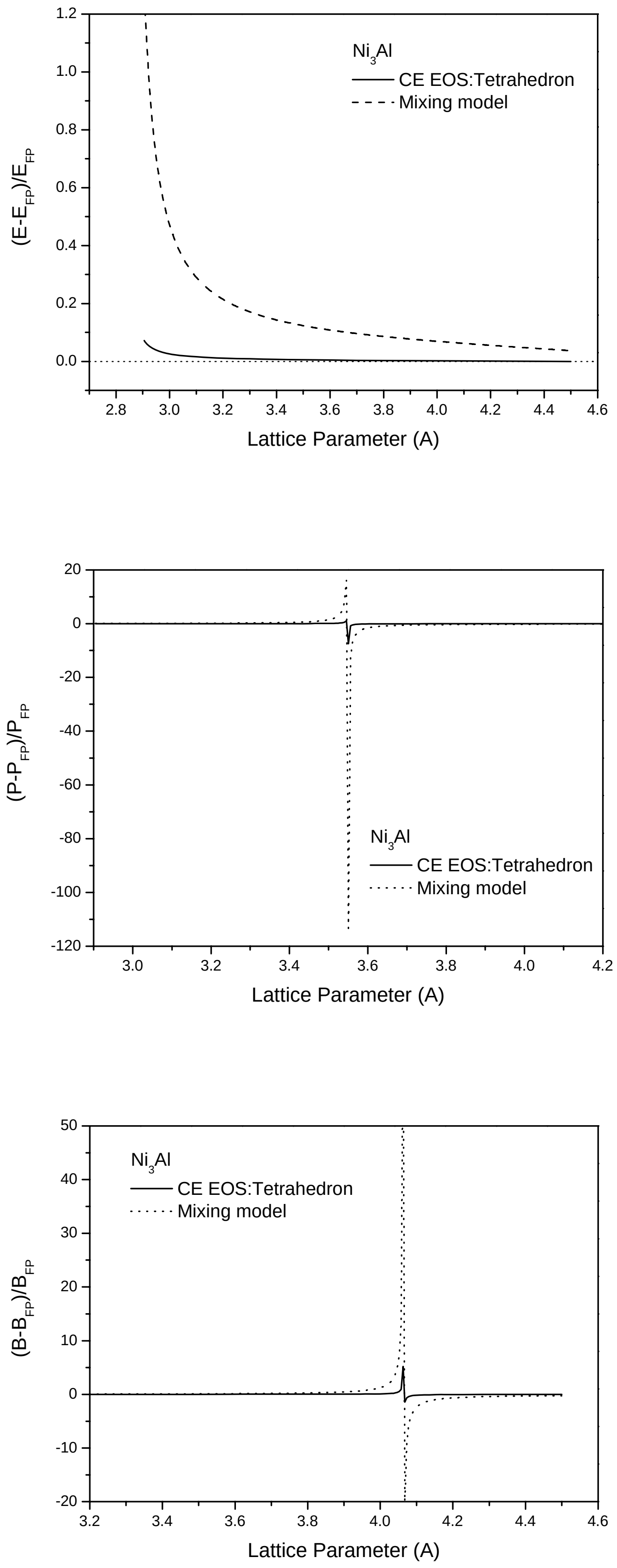

Fig. 7.

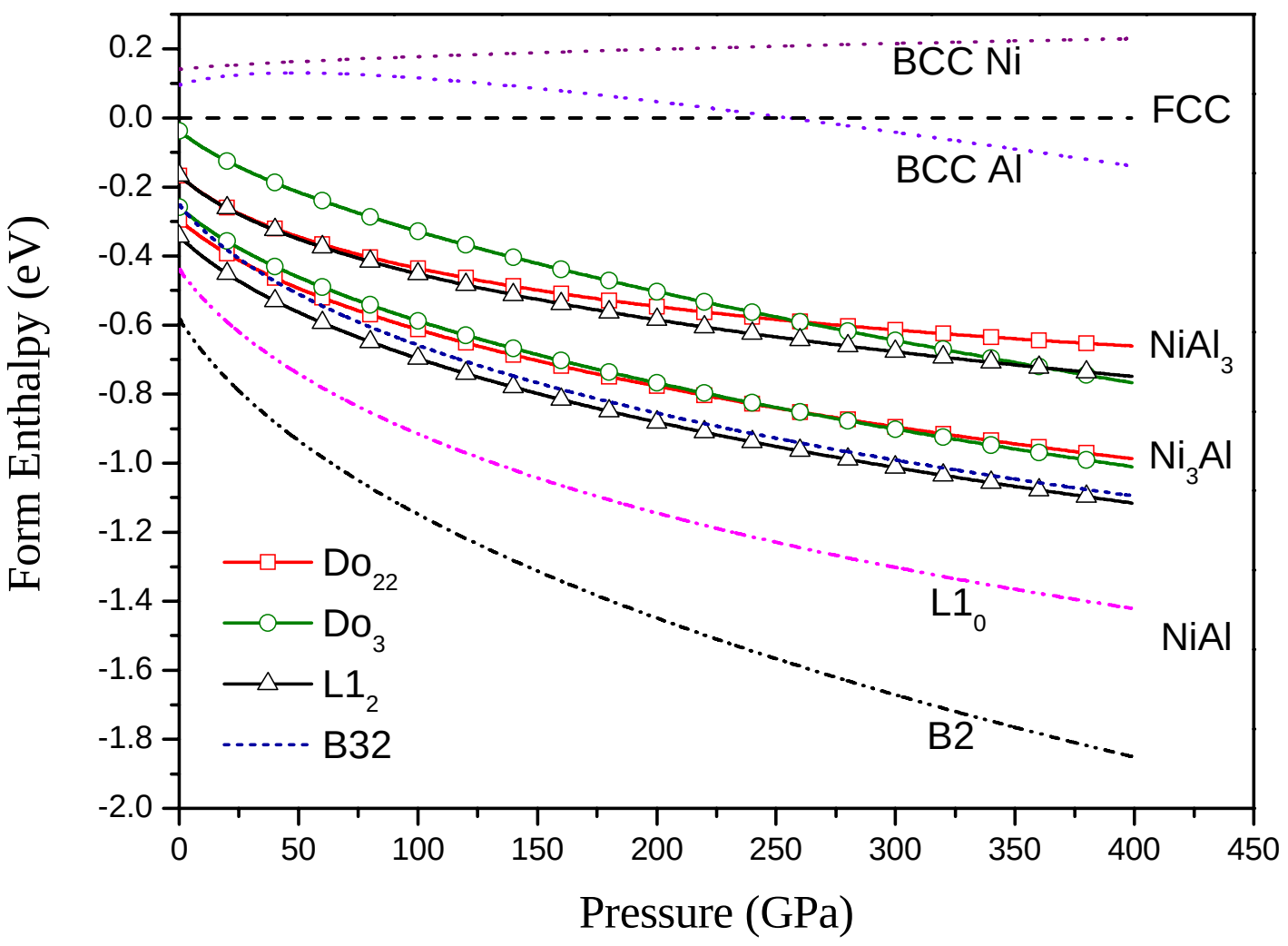

Fig.8.

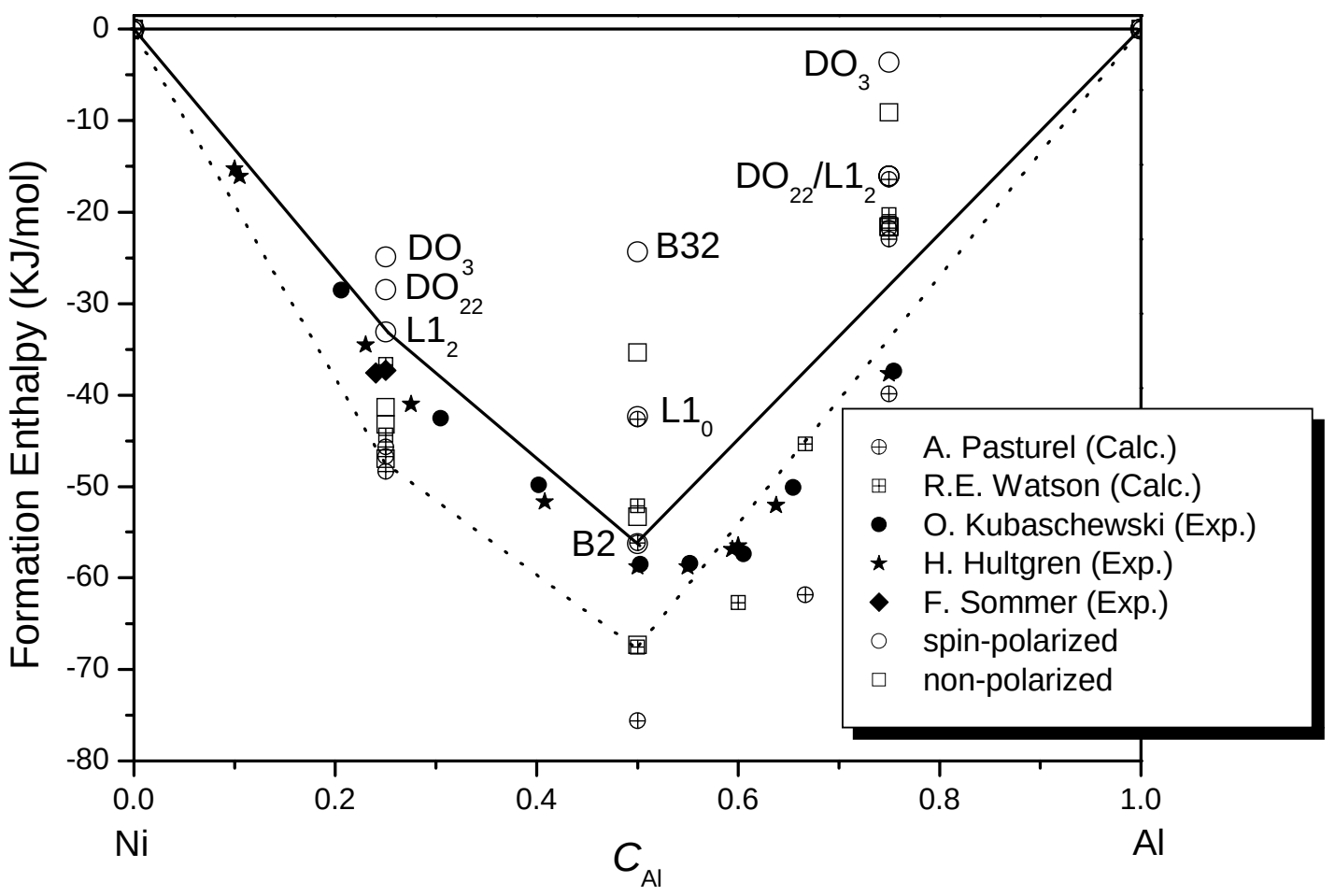

Fig.9.

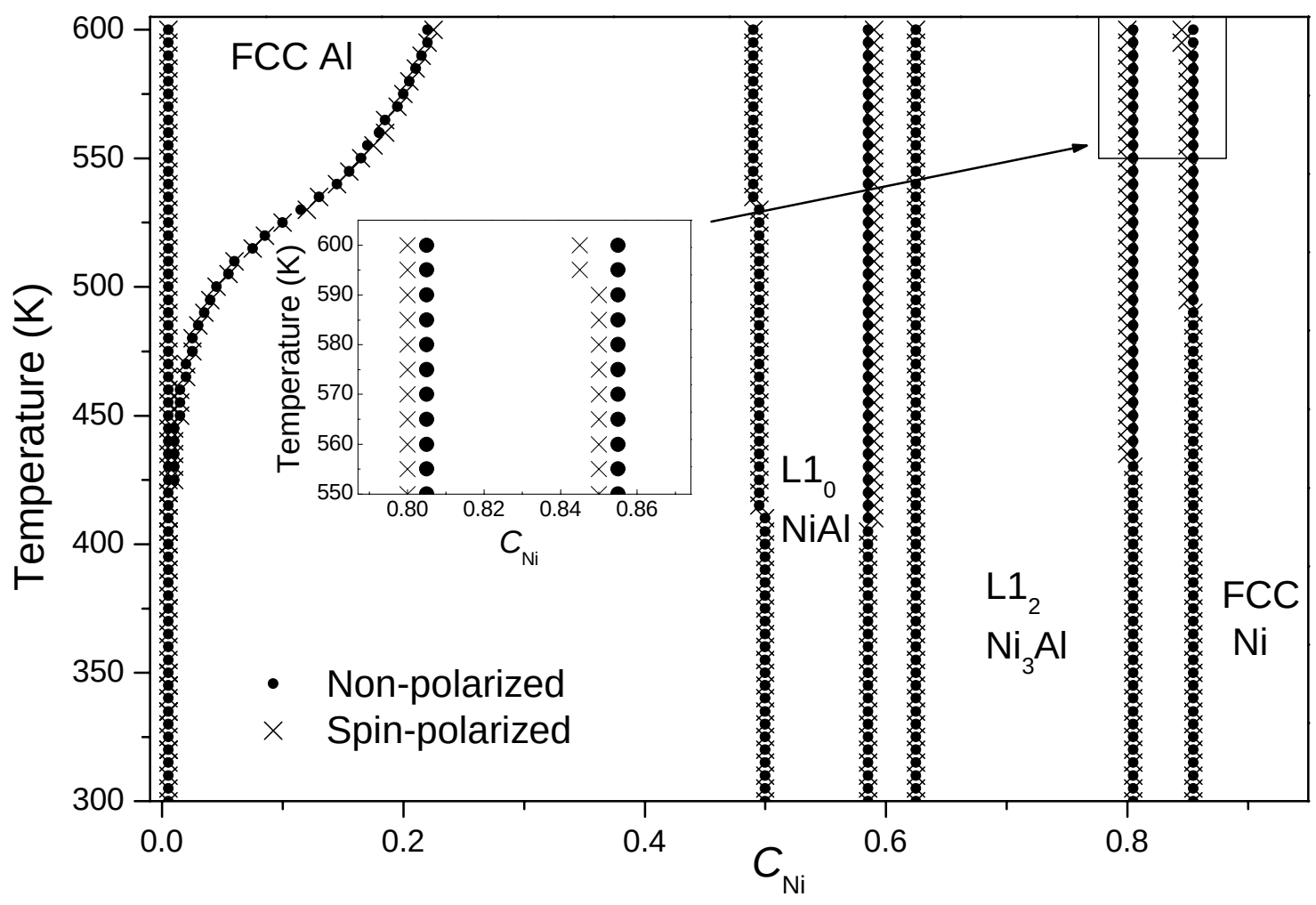

Fig. 10.

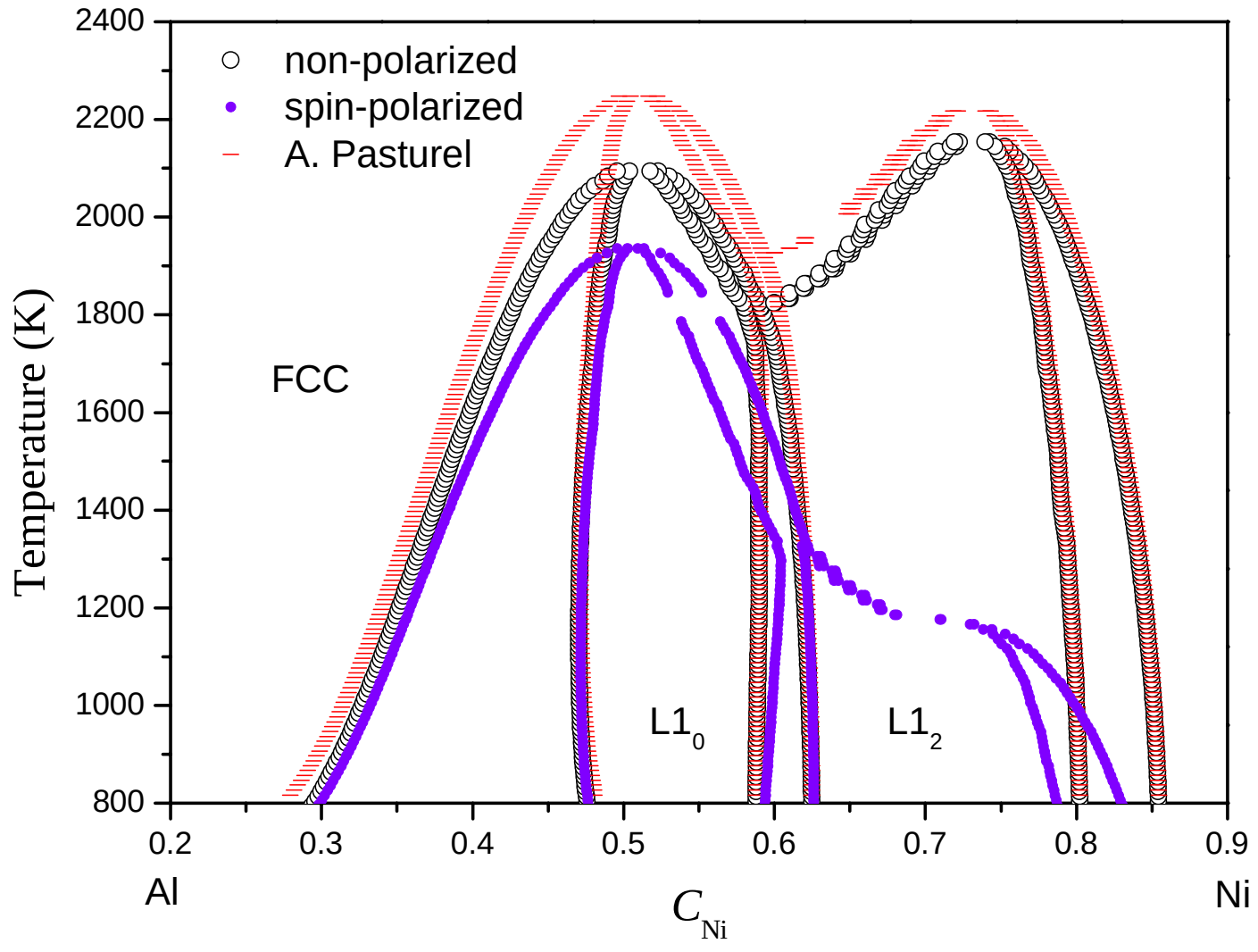

Fig. 11.

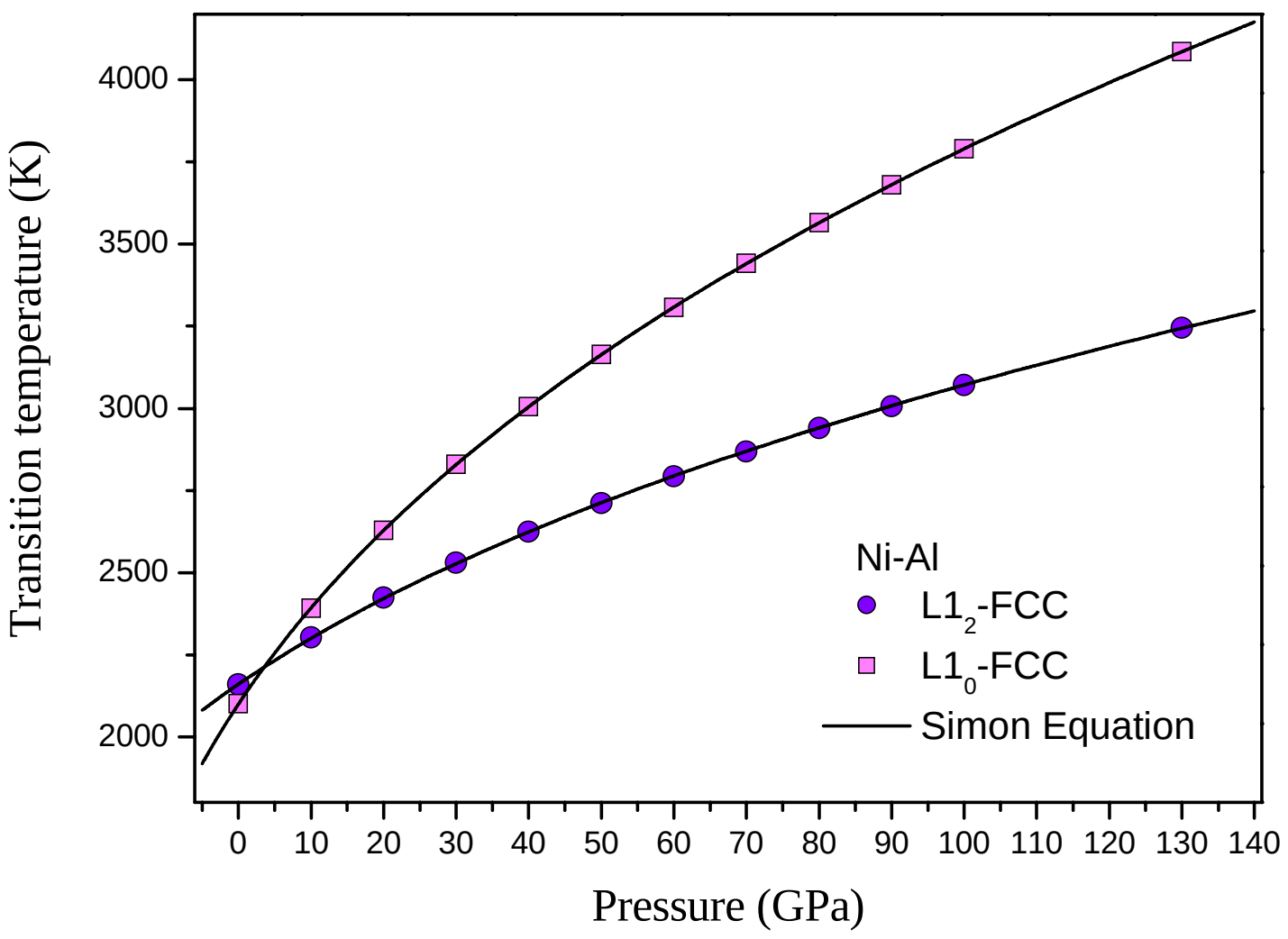
Transition temperature (K)
4000
3500
3000
2500
2000
Ni-Al
L1$_2$-FCC
L1$_0$-FCC
Simon Equation
0 10 20 30 40 50 60 70 80 90 100 110 120 130 140
Pressure (GPa)

Fig.12.